\RequirePackage{lineno}
\documentclass[aps,prl,twocolumn,superscriptaddress]{revtex4}

\usepackage{graphicx}
\usepackage{epstopdf}
\usepackage{bm}
\usepackage{amsmath}
\usepackage{amssymb}
\usepackage{hyperref}
\usepackage{float}

\begin{document}

\title{Role of higher-order phonon scattering in the zone-center optical phonon linewidth and the Lorenz oscillator model}

\author{Xiaolong Yang\footnotemark[1]} \affiliation{School of Mechanical Engineering
  and the Birck Nanotechnology Center, Purdue University, West
  Lafayette, Indiana 47907-2088, USA.} \affiliation{Institute for Advanced Study, Shenzhen University,
  Nanhai Avenue 3688, Shenzhen 518060, China}\affiliation{Frontier Institute
  of Science and Technology, and State Key Laboratory for Mechanical
  Behavior of Materials, Xi'an Jiaotong University, Xi'an 710049,
  P. R. China.}

\author{Tianli Feng\footnotemark[1]} \affiliation{School of Mechanical Engineering and
  the Birck Nanotechnology Center, Purdue University, West Lafayette,
  Indiana 47907-2088, USA.}

\author{Joon Sang Kang} \affiliation{School of Engineering and Applied Science, University of
California, Los Angeles (UCLA), Los Angeles, CA 90095, USA.}

\author{Yongjie Hu} \affiliation{School of Engineering and Applied Science, University of
California, Los Angeles (UCLA), Los Angeles, CA 90095, USA.}

\author{Ju Li} \affiliation{Department of Nuclear Science and
  Engineering and Department of Materials Science and Engineering,
  Massachusetts Institute of Technology, Cambridge, Massachusetts
  02139, USA.}

\author{Xiulin Ruan\footnotemark[2]}\affiliation{School of
  Mechanical Engineering and the Birck Nanotechnology Center, Purdue
  University, West Lafayette, Indiana 47907-2088, USA.}
\date{\today}

\begin{abstract}
Zone-center optical phonon linewidth is a key parameter for infrared and Raman spectra as well as the Lorenz oscillator model. While three-phonon scattering was often assumed to
be the leading contribution, in this work we find, surprisingly, that higher-order phonon scattering universally
plays a significant or even dominant role over three-phonon scattering at room temperature, and
more so at elevated temperatures, for a wide range of materials including diamond, Si, Ge, boron arsenide (BAs),
cubic silicon carbide (3C-SiC), and $\alpha$-quartz. This is enabled by the large four-phonon scattering phase
space of zone-center optical phonons, and distinct from heat conduction where at room temperature
four-phonon scattering is still secondary to three-phonon
scattering. Moreover, our results imply that five-phonon and even higher-order scattering may be
significant for some large band-gap materials, e.g., BAs. Our predicted infrared optical properties through the Lorenz oscillator model, after including four-phonon scattering, show much better agreement with experimental measurements than those three-phonon based predictions. This work will
raise broad interest of studying high-order scattering in various areas beyond heat conduction.
\end{abstract}

\maketitle
\renewcommand{\thefootnote}{\fnsymbol{footnote}}
\footnotetext[1]{These authors contributed equally to this work.}
\footnotetext[2]{Corresponding author. ruan@purdue.edu.}

Zone-center optical phonon linewidth, $\gamma$, is a key parameter for infrared and Raman spectra, which are basic material properties important for sensing, materials characteriations, radiative heat transfer including near-field radiation and  radiative cooling, energy harvesting, metamaterials, etc.
It is also key to the dielectric function of polar dielectrics in the infrared (IR) range, which can be effectively described by the Lorentz oscillator model\cite{Lorentz2007}:
\begin{equation}
\label{eq:1}
\epsilon(\omega)=\epsilon_{\infty}\left(1+\sum_m\frac{\omega_{\rm{LO},\it{m}}^2-\omega_{\rm{TO},\it{m}}^2}{\omega_{\rm{LO},\it{m}}^2-\omega^2-i\gamma_m\omega}  \right),
\end{equation}
where $\epsilon_{\infty}$ is dielectric constant at the high-frequency (electronic) limit, $\omega_{\rm LO}$ and $\omega_{\rm TO}$ are frequencies of zone-center TO (transverse optical) and LO (longitudinal optical) phonon modes respectively, and $m$ goes over all the IR-active zone-center phonon modes. Unfortunately, $\gamma_m$, is generally difficult to measure or predict, especially at high temperatures, $T$. Conventionally, $\gamma$ was usually obtained by fitting to the measured IR or Raman linewidths. However, due to limitations by thermal oxidation and self-radiation\cite{2007a}, most experimental measurements of $\gamma_m(T)$ cannot reach high temperatures. Moreover, the fitting approach does not allow for the predictive design of new radiative materials. Recently, first-principles methods have tackled to accurately predict \cite{Debernardi1995,Lang1999,Debernardi1998,Debernardi1999,Bechstedt1997,Tang2011}.
The predictions, however, often significantly underestimated $\gamma$ as compared to experimental data, e.g., silicon at high temperature \cite{Debernardi1995} and zinc-blende GaAs at room temperature \cite{Debernardi1998}, and the descrepancy was attributed to defect scattering, etc. An attempt to include higher-order scattering through ab initio molecular dynamics did not improve over the density-functional perturbation theory (DFPT) based on three-phonon scattering \cite{Bao2012}, hence the role of higher-order phonon scattering was unclear.

Recently, on the other hand, it was predicted that four-phonon scattering has a non-negligible role in thermal conductivity in certain materials at room temperature or high temperature\cite{Feng2016,Feng2017,Feng2018}. The prediction on BAs thermal conductivity was confirmed by experiments\cite{Kang2018,Tian2018,Li2018}. However, the higher-order scattering of zone-center optical phonons and the associated infrared and Raman properties were not particularly predicted or validated against experiments, because these phonons have negligible contribution to thermal conductivity. It should also be noted that a recent DFPT using a different four-phonon scattering method \cite{Fugallo2018} considerably overpredicted the zone-center phonon linewidth, which is difficult to interpret since the predicted intrinsic linewidth corresponds to a pristine crystal and should serve as a lower limit for experiment.

In this Letter, we clearly show that four and even higher-order phonon scattering can play a {\em dominant} role over three-phonon scattering in $\gamma_m(T)$ as well as IR properties, for a {\em wide} range of materials including GaAs, C, Si, Ge, BAs, SiC, and $\alpha$-quartz at various temperatures. In many of the situations four-phonon scattering is negligible for thermal conductivity, but important for IR and Raman properties. The phonon properties are predicted from the vibrational Hamiltonian, which can be Taylor expanded as
$\hat{H}=\hat{H}_0+\hat{H}_3+\hat{H}_4+\cdots$, where $\hat{H}_0$,
$\hat{H}_3$ and $\hat{H}_4$ are the harmonic, cubic and quartic terms,
respectively. $\hat{H}_3$ induces the phonon linewidth by three-phonon scattering $\tau_{3,\lambda}^{-1}$ \cite{Broido2007,Lindsay2010,Lindsay2012,Esfarjani2011}.
$\hat{H}_4$ induces four-phonon scattering rates $\tau_{4,\lambda}^{-1}$
\cite{Feng2016,Feng2017,Feng2018}. We also include the effect of phonon-isotope scattering $\tau_{\rm {iso}}^{-1}$ on the phonon linewidth \cite{Kundu2011}.

The full width at half maximum (FWHM) of
phonons $2\varGamma$ can be
calculated as a Matthiessen sum of contributions from isotope
scattering, 3-phonon scattering, and 4-phonon scattering:
\begin{equation}
\label{eq_solution_SMRTA}
2\varGamma = \gamma_m(T)=\tau_{\rm {iso}}^{-1}+ \tau_{3,\lambda}^{-1}+
\tau_{4,\lambda}^{-1} + ...
\end{equation}

The formalism described above requires the second, third- and fourth-order interatomic force constants (IFCs) for calculating phonon frequencies and scattering rates,
which were obtained from the VASP package\cite{Kresse1996} within the local density approximation (LDA)
with the details given in Section A of the Supplementary Material\cite{Supp}.
Our predicted phonon dispersions of all materials studied agree well with those measured in the literature as shown in Section B of the
Supplementary Material\cite{Supp}.

\begin{figure}[htp]
	\centerline{\includegraphics[width=9cm]{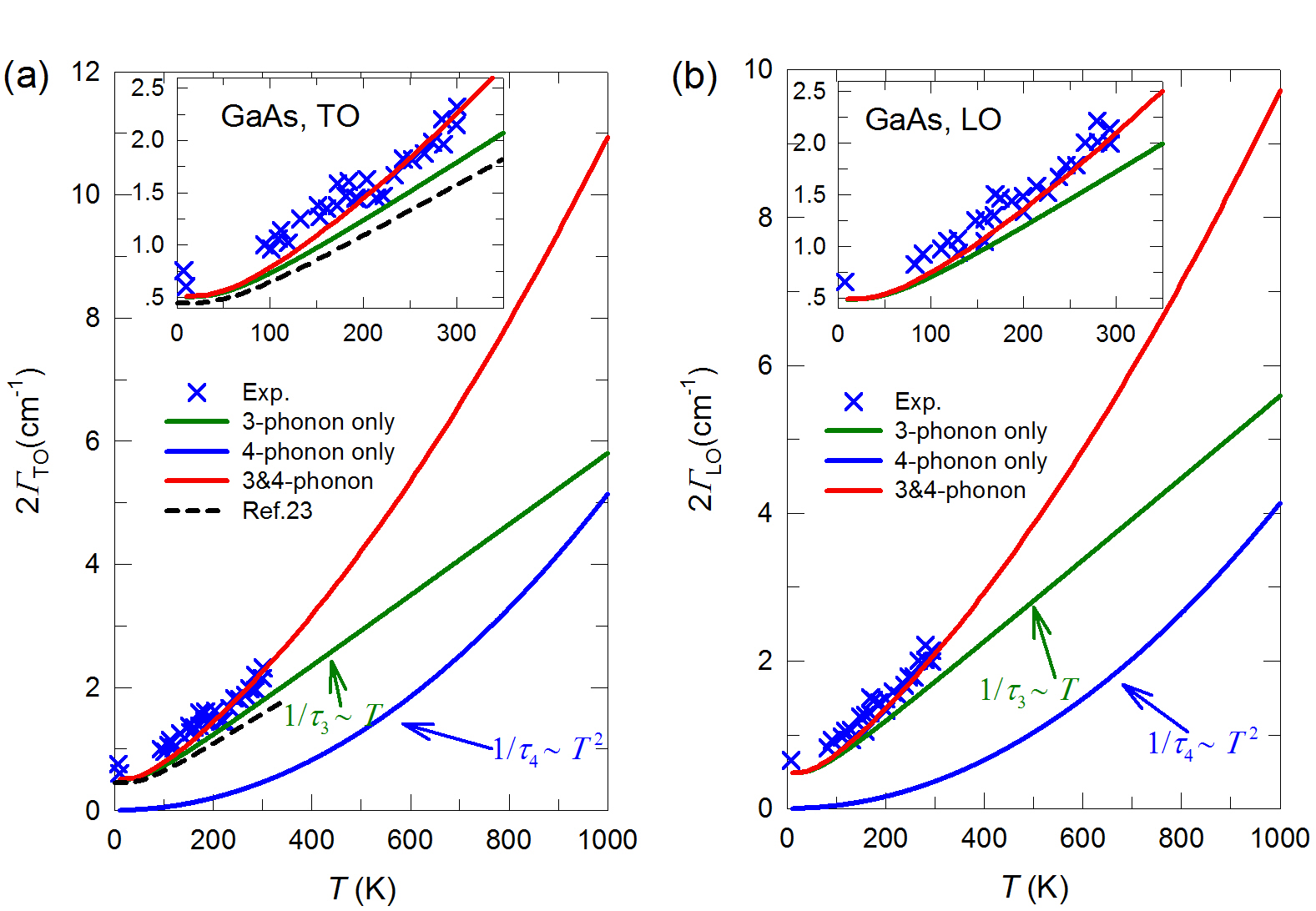}}
	\caption{\label{fig:1}
	 $T$ dependence of phonon linewidths, 2$\varGamma$, of the TO (a) and LO (b) modes in GaAs, with the insets for a better view below 400K. The solid lines represent the results of the present calculation; the dashed lines denote the results of the previous calculation from Ref.\cite{Debernardi1998}; crosses denote experimental data from \cite{Irmer1996}.}
\end{figure}

Taking GaAs as an example, we show the calculated $T$-dependent TO phonon linewidth in
Fig.\ref{fig:1}(a) with the inset for a better view below $T_{\rm
  room}=300$K. Isotope scattering is included in all curves in this paper unless noted otherwise. We find that the 3-phonon scattering prediction is considerably lower than
the experimental data \cite{Irmer1996} even at $T_{\rm room}$. With $\tau_{4,\lambda}^{-1}$ included, the prediction matches with experiment surprisingly well. Note that four-phonon scattering is negligible for thermal conductivity of GaAs at $T_{\rm room}$. Also, as $T$ increases, our prediction of 2$\varGamma$ including $\tau_{4,\lambda}^{-1}$ deviates significantly from that only
involving $\tau_{3,\lambda}^{-1}$, as shown in Fig.\ref{fig:1}(a). This indicates that 4-phonon processes, which were neglected in the past, can actually remedy the discrepancies between
previous calculations\cite{Debernardi1998} and experiments. We also note that $\tau_{3,\lambda}^{-1}$ increases approximately linearly, whereas $\tau_{4,\lambda}^{-1}$
increases approximately quadratically with increasing $T$, consistent
with Refs.\cite{Joshi1970,Feng2016}. Thus, $\tau_{4,\lambda}^{-1}$ becomes
comparable to $\tau_{3,\lambda}^{-1}$ at mid- to
high-temperatures. Similar significance of four-phonon scattering is also found for the LO
branch as shown in Fig.\ref{fig:1}(b).

\begin{figure*}[htp]
	\centerline{\includegraphics[width=13cm]{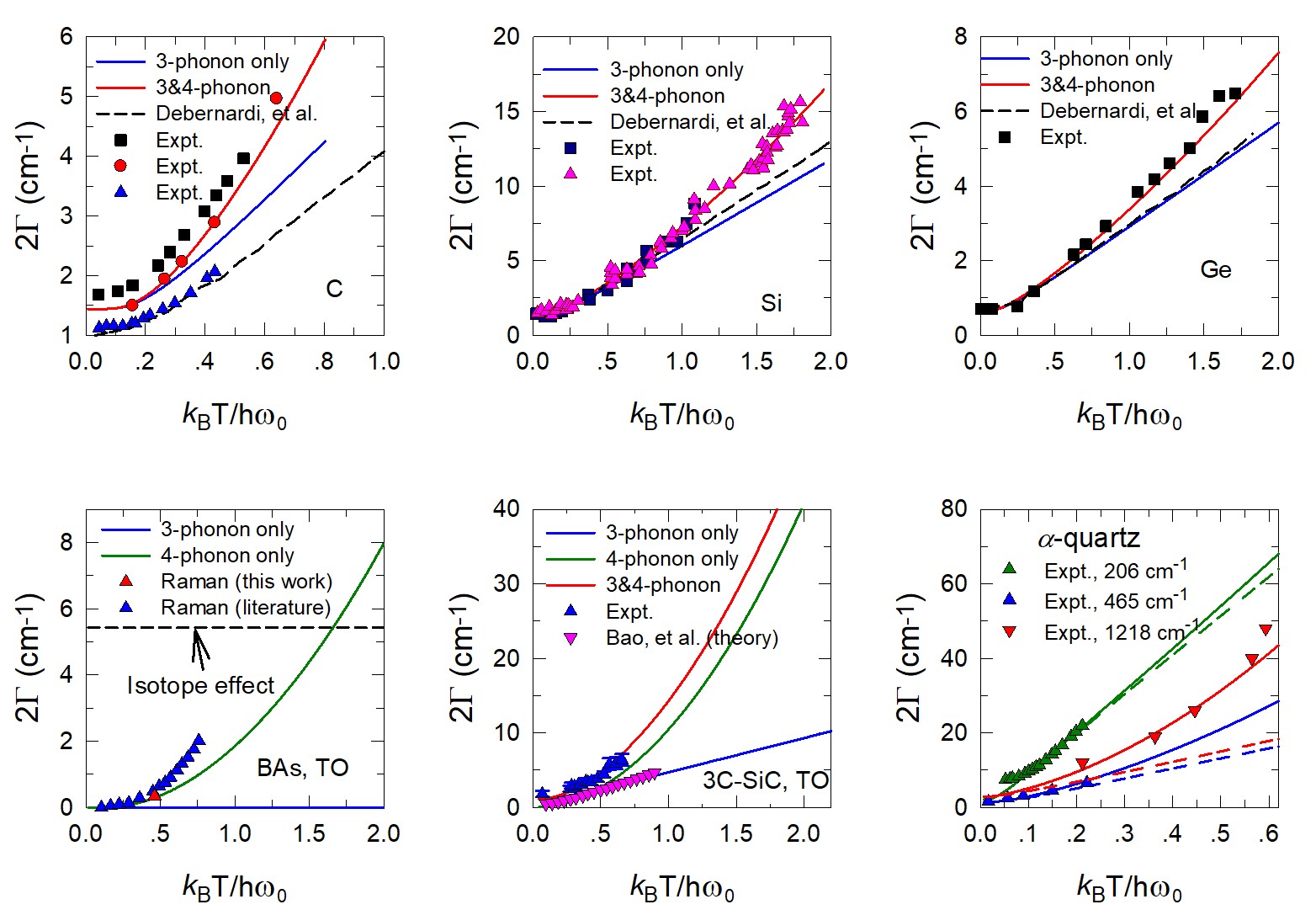}}
	\caption{\label{fig:2} Zone-center optical phonon linewidth in C, Si, Ge, BAs, 3C-SiC, and $\alpha$-quartz as a function of the normalized temperature $k_{\rm B}T/h\omega_{\rm0}$, $\omega_{\rm0}$ being the Raman phonon frequency. For C, Si and Ge, the solid curves represent the results of the present calculation, and the dashed curves denote the results of the previous calculation from Ref.\cite{Debernardi1998}; all symbols denote experimental data from Refs.\cite{Borer1971,Herchen1991,Balkanski1983,Menendez1984}. For BAs, the solid and dashed curves represent the results of
		our present calculations on the TO phonon linewidth; the red up triangle denotes our Raman measurement result; blue up triangles denote
		experimental data from Ref.\cite{Hadjiev2014}.
		For 3C-SiC, the solid curves represent our present prediction of TO phonon linewidth; blue up triangles denote experimental data
		from Ref.\cite{Ulrich1999}; pink down triangles from
		Ref.\cite{Tong2018}. For $\alpha$-quartz, all curves represent our predicted phonon linewidth with
		(solid lines) and without (dashed
		lines)$\tau_{4,\lambda}^{-1}$ for phonon modes with frequencies of 206 $\rm cm^{-1}$, 465 $\rm cm^{-1}$, and 1218 $\rm cm^{-1}$, and all symbols denote the
		experimental data from Refs.\cite{Dean1982,Gervais1975}.}
\end{figure*}

Fig.\ref{fig:2} shows the $T$-dependent optical phonon linewidths for C, Si, Ge, BAs,
3C-SiC, and $\alpha$-quartz, respectively. Similarly, only after including 4-phonon scattering we can see good agreements with available experimental data from Refs.\cite{Borer1971,Herchen1991,Menendez1984,Hadjiev2014,Ulrich1999,Dean1982,Gervais1975} and our own Raman spectra of BAs (see Section C of the Supplementary Material\cite{Supp} for details). It should be noted that unlike other materials, the experimental Raman spectra of BAs contain strong influence from isotope or defect scattering. To highlight only the $T$-dependent anharmonic effects, we have subtracted such background contribution from both the experimental and theoretical data for comparison. It can be seen that in BAs the 3-phonon processes have little contribution to the zone-center TO phonon linewidth due to lack of allowed scattering, and the 4-phonon scattering totally dominates over three-phonon scattering in the entire $T$ range\cite{Feng2017}. This is unlike the room-temperature thermal conductivity of BAs, where the role of four-phonon scattering, although significant, is still less important than three-phonon scattering. Remarkably, in BAs phonon-isotope scattering is particularly important and can play a dominant role over the intrinsic phonon-phonon scattering even at $T_{\rm room}$ and higher ($k_{\rm B}T/h \omega_{\rm 0} >$  0.45). Also note that for BAs at higher temperature, our prediction is still quite smaller than experiment even after including $\tau_{4,\lambda}^{-1}$, implying that even higher-order such as five-phonon scattering cannot be neglected in this particlar material for its IR properties. It needs to mention that the discrepancy between our calculation and experiment also suggests possible contributions to the scattering entropies from the $T$-induced phonon renormalization\cite{Hellman2013}, which are not dealt with in the present work, but non-negligible especially above Debye temperature ($k_{\rm B}T/h \omega_{\rm 0} >$  1).

\begin{figure}[htp]
	\centerline{\includegraphics[width=9cm]{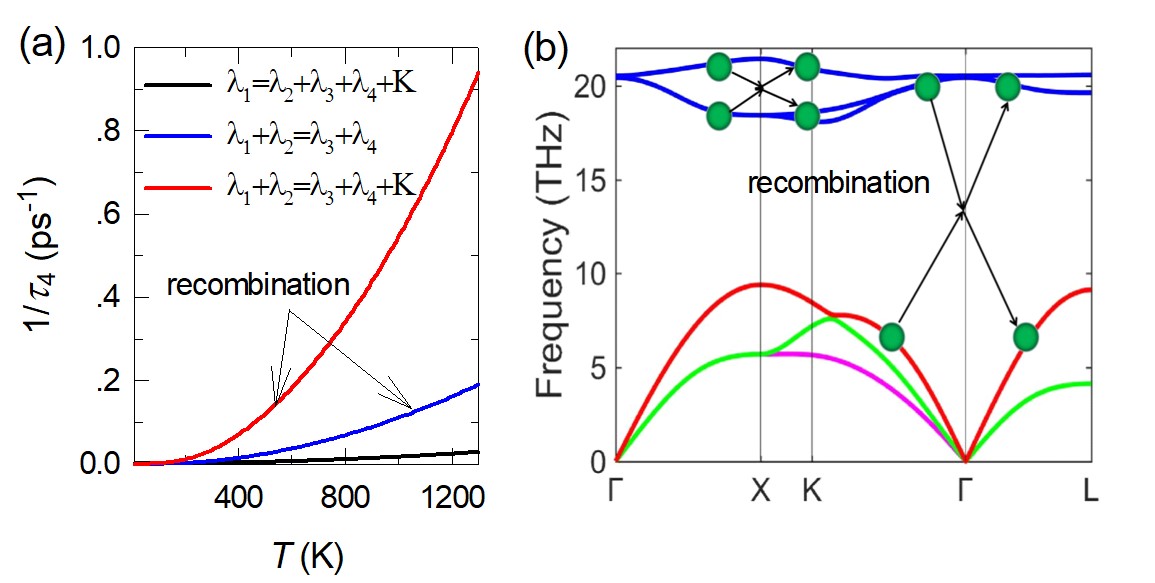}}
	\caption{\label{fig:3} (a) The contribution to $\tau_{4,\lambda}^{-1}$ of BAs from different scattering channels.
	(b) Phonon dispersion of BAs with schematic diagrams of four-phonon scattering processes dominated by recombination channels.}
\end{figure}

To gain a deeper insight into the microscopic mechanisms that govern the four-phonon decay process for optical modes, taking BAs as an example, we further analyze the contribution to $\tau_{4,\lambda}^{-1}$ from different scattering channels. From Fig.\ref{fig:3}(a), it can be seen that the dominant decay channels for zone-center optical phonons are recombination processes including momentum-conserved Normal processes (\textbf{K}=0) $\rm \lambda_{1}+\lambda_{2} \longrightarrow \lambda_{3}+\lambda_{4}$, and Umklapp processes ($\textbf{K} \neq 0$) $\rm \lambda_{1}+\lambda_{2} \longrightarrow \lambda_{3}+\lambda_{4}+\textbf{K}$, which contribute more than 90$\%$ to $\tau_{4,\lambda}^{-1}$ at 1000K. Similar cases are also found in the other materials studied. We thus conclude that in all these materials, the high four-phonon scattering rates $\tau_{4,\lambda}^{-1}$ of optical branches mainly stem from the large phase space of the recombination process $\rm \lambda_{1}+\lambda_{2} \longrightarrow \lambda_{3}+\lambda_{4}$. The origin of such decay processes in BAs is illustrated in Fig.\ref{fig:3}(b). Due to the bunching of optical branches, recombination processes can easily satisfy the energy and momentum conservation rules compared to three-phonon scattering. This is a key difference compared to the four-phonon scattering of acoustic branches which are important for thermal conductivity, which strongly depends on the temperature, a-o gap, and anharmonicity of the materials. Therefore, the significance of four-phonon scattering will be more broadly seen in zone-center optical phonon linewidth than in thermal conductivity.

\begin{figure*}[htp]
	\centerline{\includegraphics[width=15cm]{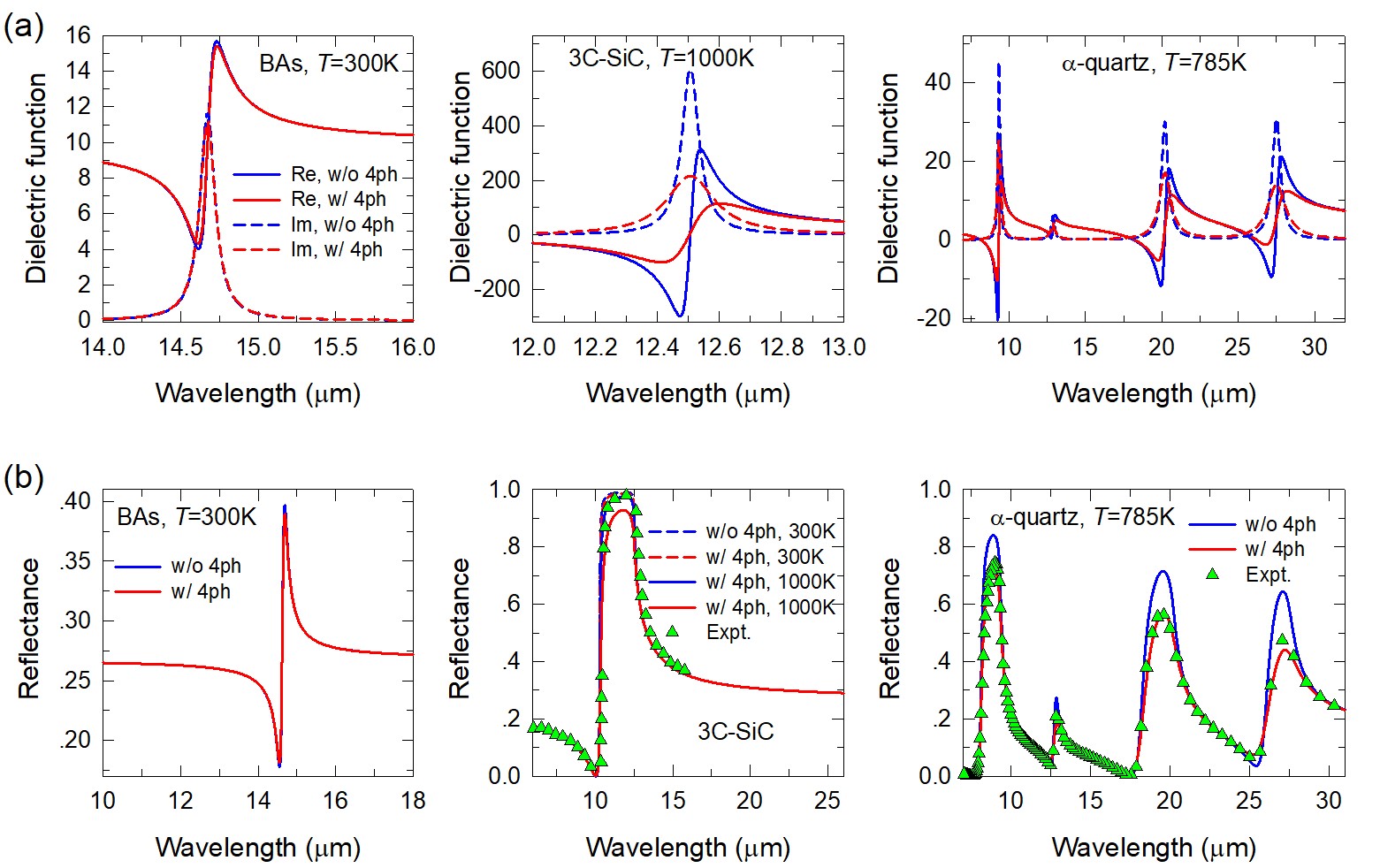}}
	\caption{\label{fig:4} (a) The calculated dielectric function of BAs,
          3C-SiC, and $\alpha$-quartz with (red lines) and without
          (blue lines) $\tau_{4,\lambda}^{-1}$ included. Solid lines
          represent the real part (Re) of dielectric function
          $\epsilon (\omega)$ and dashed lines denote the imaginary
          part (Im) of dielectric function $\epsilon$. (b) The calculated semi-infinite normal reflectance $R(\omega)$ for BAs at 300K, SiC at 300K (dashed curves) and 1000K (solid curves), and $\alpha$-quartz at 785K along with available experimental data \cite{Pitman2008,Gervais1975} for comparison. }
\end{figure*}

With the predicted 2$\varGamma(T)$ combined with frequency values
$\omega_{\rm TO}$ and $\omega_{\rm LO}$ listed in Supplemental Table S1 \cite{Supp}, we
address the importance of four-phonon scattering in dielectric
function. The calculated $\epsilon(\omega)$ data with
$\tau_{4,\lambda}^{-1}$ (red lines) are compared with those without
$\tau_{4,\lambda}^{-1}$ (blue lines) for BAs, 3C-SiC and $\alpha$-quartz at various $T$,
as shown in Fig.\ref{fig:4}(a). In BAs, although four-phonon process is extremely significant even at $T_{\rm room}$, however, it does not have obvious impact on the dielectric function at $T_{\rm room}$, owing to the exceptional strong phonon-isotope scattering as mentioned above. For isotopically pure BAs, the four-phonon scattering will make a large difference. For 3C-SiC, with $\tau_{4,\lambda}^{-1}$ included, the imaginary part of
the dielectric function, $\epsilon_{i}(\omega)$, has a much broader peak,
corresponding to a larger linewidth as highlighted by the solid
arrow. We also find that when four-phonon scattering is added the peak value of $\epsilon_{i}(\omega)$
decreases by 70$\%$ for 3C-SiC at 1000K. For $\alpha$-quartz, after
introducing $\tau_{4,\lambda}^{-1}$, the peak values of the dielectric
function at 785K also have a significant reduction ranging from 20$\%$
to 50$\%$ for different vibrational modes. These results suggest that
previous works that considered three-phonon anharmonicity alone led to significant errors,
and four-phonon scattering plays a decisive role in
predicting high-temperature dielectric functions of
IR-active materials.

With the complex dielectric function $\epsilon (\omega)$, normal
reflectance from a semi-infinite block $R$ can be calculated as
$R(\omega)=\left|\frac{\sqrt{\epsilon(\omega)-1}}{\sqrt{\epsilon(\omega)+1}}\right|^2$.
In Fig.\ref{fig:4}(b), we show our predicted reflectance $R(\omega)$ of BAs, 3C-SiC, and $\alpha$- quartz at various $T$, as compared to the available experimental measurement for $\alpha$- quartz in Ref.\cite{Gervais1975}. It is found that at $T_{\rm room}$ four-phonon scattering cannot affect the reflectance of BAs significantly, and the peak of BAs reflectivity in range of 14-15 $\mu$m has only a slight drop when four-phonon scattering is included. For 3C-SiC, with four-phonon scattering included, the peak of reflectivity in range of 10 - 13 $\mu$m will reduce by 10$\%$. For $\alpha$-quartz, it can be observed that the predicted reflectance peak with $\tau_{4,\lambda}^{-1}$ decreases by 10 $\sim$ 30$\%$ as compared to that with only $\tau_{3,\lambda}^{-1}$, and matches exceptionally well with the existing IR data \cite{Gervais1975}.

Moreover, these zone-center phonon modes would have extremely small or even zero mean free path (MFP) due to their small group velocity and high scattering rates. Recently, a nice two-channel thermal conductivity model has pointed out that these modes with MFP smaller than the Ioffe-Regel limit can no longer be treated as phonons but a hopping channel should be introduced \cite{Mukhopadhyay2018}. Our work, through matching the phonon linewidth with experiment, indicates that these modes are still well defined phonon modes with scattering rates correctly described by first principles. This suggests that the two-channel concept of Ref.\cite{Mukhopadhyay2018} may be modified in the way that these modes are still treated as phonons, but their heat conduction should be described by a different theory than the BTE.

In conclusion, we have established a significant or even dominant role of four-phonon scattering and even higher-order scattering in determining the zone-center optical phonon linewidth for a series of important materials, using first principle-based DFT with perturbation theory. Our results demonstrate that the previously neglected four-phonon scattering generally dominates optical phonon linewidth at elevated and high temperatures, due to the highly restrictive selection rules of three-phonon scattering that can hardly be met in these materials. With four-phonon scattering included, the infrared optical properties of $\alpha$-quartz agree well with previous measurements. Our work thus reveals a more crucial role of higher-order phonon scattering in IR optical properties than in heat conduction.


\end{document}